\documentclass[acmtog]{acmart}

\usepackage[capitalize]{cleveref}
\usepackage{multirow}
\usepackage{enumitem}
\usepackage{makecell}
\usepackage[ruled]{algorithm2e} 
\newcommand{\ours}{CoherentRaster}

\newcommand{\viewreuse}{Cross-view Coherent Attribute Reuse}
\newcommand{\viewremapping}{View-coherent Remapping}

\AtBeginDocument{%
  }

\copyrightyear{2026}
\acmYear{2026}
\setcopyright{cc}
\setcctype{by}
\acmConference[SIGGRAPH Conference Papers '26]{Special Interest Group on Computer Graphics and Interactive Techniques Conference Conference Papers}{July 19--23, 2026}{Los Angeles, CA, USA}
\acmBooktitle{Special Interest Group on Computer Graphics and Interactive Techniques Conference Conference Papers (SIGGRAPH Conference Papers '26), July 19--23, 2026, Los Angeles, CA, USA}
\acmDOI{10.1145/3799902.3811217}
\acmISBN{979-8-4007-2554-8/2026/07}

\begin{document}
\title{\ours{}: Efficient 3D Gaussian Splatting for Light Field Displays}

\author{Gyujin Sim}
\affiliation{
  \institution{POSTECH}
  \city{Pohang}
  \country{Republic of Korea}
}
\email{sgj0402@postech.ac.kr}

\author{Seungjoo Shin}
\affiliation{
  \institution{POSTECH}
  \city{Pohang}
  \country{Republic of Korea}
}
\email{seungjoo.shin@postech.ac.kr}

\author{Hosung Jeon}
\affiliation{
  \institution{ETRI}
  \city{Daejeon}
  \country{Republic of Korea}
}
\email{h.jeon@etri.re.kr}

\author{Gwangsoon Lee}
\affiliation{
  \institution{ETRI}
  \city{Daejeon}
  \country{Republic of Korea}
}
\email{gslee@etri.re.kr}

\author{Hyon-Gon Choo}
\affiliation{
  \institution{ETRI}
  \city{Daejeon}
  \country{Republic of Korea}
}
\email{hyongonchoo@etri.re.kr}

\author{Sunghyun Cho}
\affiliation{
  \institution{POSTECH}
  \city{Pohang}
  \country{Republic of Korea}
}
\email{s.cho@postech.ac.kr}

\begin{abstract}

Light field displays (LFDs) require rendering an interlaced image that encodes many view‑dependent observations. This multi‑view requirement introduces substantial computational overhead, making real‑time rendering difficult to achieve.
While 3D Gaussian Splatting (3DGS) is efficient for single-view rendering on 2D displays, directly extending it to LFDs is computationally expensive.
Moreover, prior accelerations either suffer from GPU inefficiency under spatially incoherent subpixel layouts or rely on computationally heavy multi-plane intermediates.
In this paper, we propose \textit{CoherentRaster}, a 3DGS-based light field rendering framework that performs subpixel-level rasterization.
Our method employs \viewreuse{} to eliminate redundant computation across neighboring viewpoints and applies \viewremapping{} to restore warp-level memory efficiency degraded by the interlaced subpixel layout.
Together, CoherentRaster provides an efficient pipeline for real-time, high-quality light field synthesis on consumer-grade hardware.
The code is at https://github.com/sgj0402/coherent-raster.
\end{abstract}

\ccsdesc[500]{Computing methodologies~Rendering}

\keywords{3D Gaussian Splatting, Light Field Display, Rasterization}


\begin{teaserfigure}
  \centering 
    \includegraphics[width=0.9\textwidth]{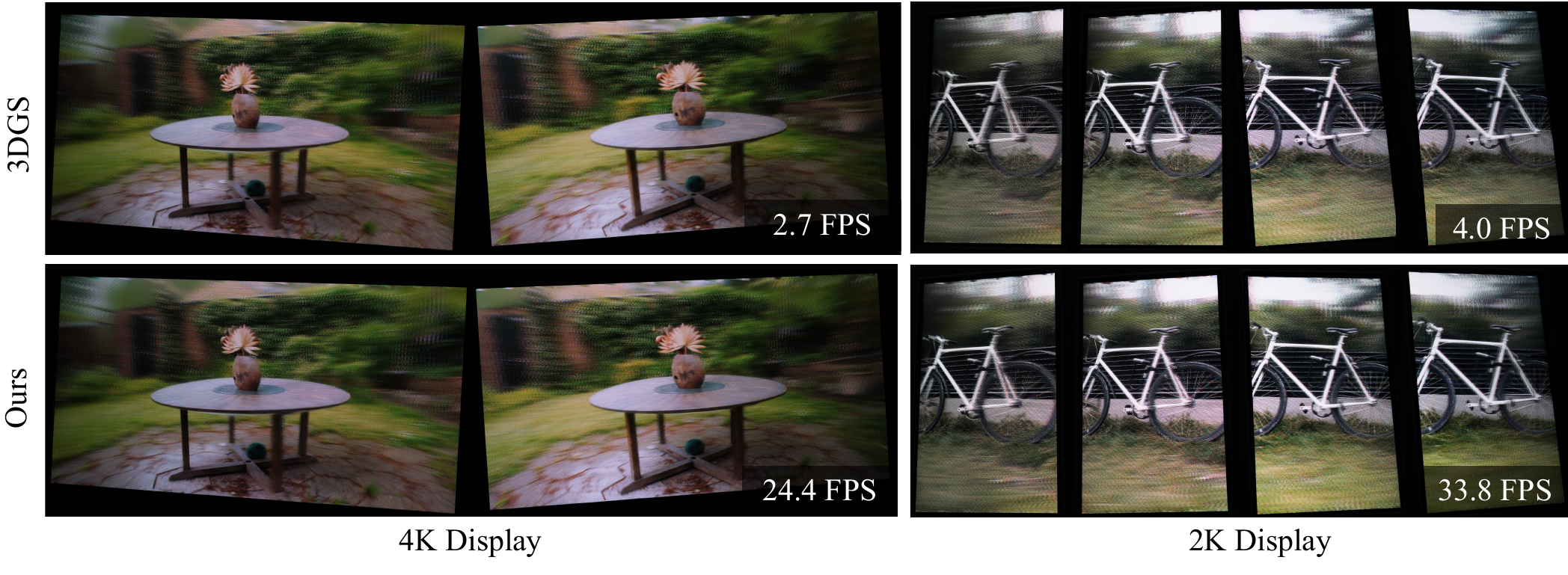}
    \caption{\textbf{Through-the-lens comparison of rendered light field image.} We visualize the rendered content on a physical Light Field Display, captured from left and right viewpoints. Comparing Full-frame rendering 3DGS and \ours{}, our method achieves significantly higher frame rates (FPS) while maintaining visual quality. (\textit{Garden} and \textit{Bicycle} scenes from the Mip-NeRF 360 dataset)
    }
    \Description{teaser}
    \label{fig:teaser}
\end{teaserfigure}

\maketitle

\section{Introduction}
\label{sec:introduction}

Light field displays (LFDs) provide glasses-free autostereoscopic 3D visualization with continuous motion parallax, enabling immersive viewing experiences for 3D content without wearable devices. Unlike conventional single-view 2D displays, LFDs render interlaced light field images to simultaneously deliver view-dependent visual outcomes across a finite range of viewing angles. As LFDs with high spatial and angular resolution have recently become commercially available, there exists a growing demand for high-quality 3D content tailored to these 3D displays~\cite{looking_glass_factory,leia,sony}.

Recent advances in radiance field representation~\cite{nerf,3dgs} have opened up their potential to serve as effective 3D content for LFDs. Specifically, their ability to represent complex 3D scenes and support high-quality view synthesis significantly lowers the barrier to 3D content creation for everyday users. Among them, 3D Gaussian Splatting (3DGS)~\cite{3dgs} has emerged as a practical solution for real-time, high-quality novel viewpoint rendering, achieving hundreds of frame rates on conventional 2D displays through efficient tile-based rasterization.
However, applying 3DGS directly to light field displays is challenging, as LFDs require simultaneously synthesizing many view-dependent images rather than a single viewpoint. This fundamental difference introduces new computational bottlenecks that conventional 3DGS pipelines are not designed to handle.

To enable real‑time, high‑resolution 3DGS rendering on LFDs, an LFD‑specialized rasterization pipeline is required. Traditional light field reconstruction pipelines render full images for all target viewpoints and then interlace them into a single light field image~\cite{qi2022dense,chen2019dense}. However, synthesizing these dense viewpoints is extremely costly: the computation and memory usage grow linearly with the number of required views, and the high resolution of interlaced images further amplifies the burden~\cite{lf_display_and_rendering_survey}. In practice, LFDs often require tens to hundreds of viewpoints at 2K or higher resolution, making na\"ive multi‑view rendering prohibitively slow and preventing interactive performance. These challenges underscore the need for rendering strategies tailored specifically to 3D displays.

To accelerate light field rendering, DirectL~\cite{directl} first introduced subpixel-level rendering, leveraging the observation that an interlaced light field image uses only a small subset of subpixels from different viewpoints, as shown in \cref{fig:lenticular_lfd_viewpoint_number_matrix}.
Instead of contributing all pixels from all views, the final image is formed by interlacing only these selected subpixels. Consequently, rendering full images for every viewpoint performs substantial computation on pixels that never appear in the output. DirectL avoids this wasted work by sampling rays only for the subpixels that actually contribute to the interlaced image, bypassing full multi‑view rendering. Building on this idea, \citet{text_driven_lf_editing_gs} extend subpixel‑level rendering to 3DGS by evaluating only the Gaussians required for the visible subpixels.
However, applying this strategy to 3DGS introduces a new challenge: because adjacent subpixels often originate from different viewpoints, the GPU loses the \emph{spatial coherence} it typically relies on. Modern GPUs execute threads in lockstep groups called warps (32 threads), making coherent memory access crucial for high throughput. When neighboring threads access disjoint view-dependent Gaussian attributes, warp‑level efficiency degrades significantly.

Another promising direction to efficient light field rendering would be to leverage \emph{cross‑view coherence}, which refers to the fact that neighboring viewpoints often observe similar scene content. Specifically, by exploiting the concept of multi‑plane images, Kim et al.~\shortcite{g2lf} introduce depth‑aligned plane representations that allow neighboring viewpoints to share intermediate rendering results. However, this formulation faces a critical scalability bottleneck for high‑resolution LFDs. Achieving high‑fidelity rendering often requires hundreds of planes to suppress depth discretization artifacts, an overhead that grows rapidly with spatial resolution. As we further analyze in the experimental section, this dense intermediate representation leads to substantial computational costs, making real‑time rendering impractical on consumer‑grade hardware.

In this paper, we introduce \textit{\ours{}}, a real‑time 3DGS‑based light field rendering framework that directly addresses the key limitations of prior approaches. First, we adopt subpixel‑level rasterization to evaluate only the subpixels that contribute to the interlaced light field image, eliminating the unnecessary work inherent in full multi‑view rendering. Second, we exploit cross‑view coherence by reusing Gaussian attributes that vary smoothly across views, substantially reducing per‑view computation.
Finally, we resolve the GPU inefficiency caused by the spatially incoherent subpixel layout through a view‑coherent remapping strategy that reorganizes threads to improve memory coalescing and warp‑level execution. Together, these components form a lightweight and scalable pipeline that leverages the benefits of cross‑view coherence, similar in spirit to MPI‑based methods, while avoiding the heavy intermediate representations required by multi‑plane approaches, enabling real‑time light field synthesis on commodity GPUs.

We evaluate \ours{} on both synthetic and real-world benchmarks adapted for light field rendering.
\ours{} achieves up to 23 FPS for 4K ($3840 \times 2160$) interlaced outputs for real-world 3D scenes with 71 viewpoints on an RTX 5090 GPU.
This demonstrates real-time rendering capabilities while maintaining comparable visual quality, as shown in \cref{fig:teaser}.

Our contributions are summarized as follows:
\begin{itemize} [leftmargin=*]
    \item We introduce \ours{}, an efficient 3DGS-based light field rendering framework that enables real-time, high-resolution 3D visualization on LFDs.
    \item We propose \viewreuse{} that shares cross-view Gaussian evaluations to reduce redundant computation across adjacent viewpoints.
    \item We present \viewremapping{} that reorders thread-to-subpixel mapping to restore warp-level memory efficiency under interlaced subpixel layouts.
\end{itemize}

\section{Related Work}
\label{sec:related_work}



\subsection{Novel View Synthesis}

Novel view synthesis aims to render a scene from arbitrary camera positions using pre-captured images. Early image-based rendering methods~\cite{levoy1996lightfieldrendering, gortler1996lumigraph} achieved this by resampling a 4D radiance function constructed from dense image arrays. An alternative approach bypasses this extensive capture by augmenting a reference image with depth. DIBR~\cite{fehn2004depth} warps a single color image into a target viewpoint using its per-pixel depth map, while layered depth images~\cite{shade1998layered} store multiple depth samples along each ray to capture surfaces hidden behind the foreground, naturally resolving disocclusions during warping. As another depth-based representation, Multiplane images (MPIs)~\cite{zhou2018stereo} represent a scene as a stack of fronto-parallel RGBA planes composited via alpha blending. More recently, NeRF~\cite{nerf} and 3DGS~\cite{3dgs} have demonstrated remarkable success in reconstructing complex 3D scenes from captured images, making them a compelling foundation for 3D content on light field displays.

\subsection{3DGS Acceleration for Single-View Rendering}

3DGS~\cite{3dgs} achieves real-time view synthesis through
differentiable tile-based rasterization, and numerous follow-up works have focused
on further accelerating this pipeline. One line of research reduces the number of
Gaussian primitives through pruning strategies guided by rendering contribution
\cite{fang2024mini,niemeyer2025radsplat}, spatial occupancy~\cite{fan2024lightgaussian},
or learnable masks~\cite{lee2024compact}. Another direction improves rasterization
efficiency by tightening projected Gaussian bounds, thereby reducing sorting and
blending overhead without compromising fidelity~\cite{adr_gaussian,speedy_splat}.

While these techniques substantially improve single-view performance, they do not
address the multi-view scalability required for light field displays, where tens to
hundreds of viewpoints must be synthesized for each frame. Our work targets this
multi-view setting by designing an LFD-tailored 3DGS pipeline that remains efficient
even under dense viewpoint requirements.

\subsection{Light Field Rendering}

Light field displays require rendering many view‑dependent images, motivating techniques that reduce redundant computation across viewpoints.
DirectL~\cite{directl} introduced subpixel‑level rendering, which avoids evaluating unused pixels by rendering only the subpixels that contribute to the interlaced output. \citet{text_driven_lf_editing_gs} applied this idea to 3DGS, enabling efficient light field synthesis. However, subpixel‑level rendering disrupts spatial locality in rasterization‑based pipelines, causing neighboring threads to access disjoint view-dependent Gaussian attributes and degrading memory coalescing and warp‑level efficiency. Our method resolves this issue through a view‑coherent remapping strategy that restores GPU execution coherence under subpixel‑level rendering.

In parallel, \citet{g2lf} introduce an MPI-based representation designed to leverage cross-view coherence. Specifically, adjacent viewpoints share intermediate rendering results in the multiple planes, thereby reducing redundant computation among viewpoints. However, such a representation remains a fundamental scalability bottleneck that arises from the inherent trade-off between plane resolution and visual fidelity. Conversely, our method shares intermediate results across neighboring viewpoints by reusing precomputed Gaussian attributes, which avoids the need for auxiliary representations and prevents discretization errors tied to their resolution.

\begin{figure}[t]
    \centering 
    \includegraphics[width=0.9\linewidth]{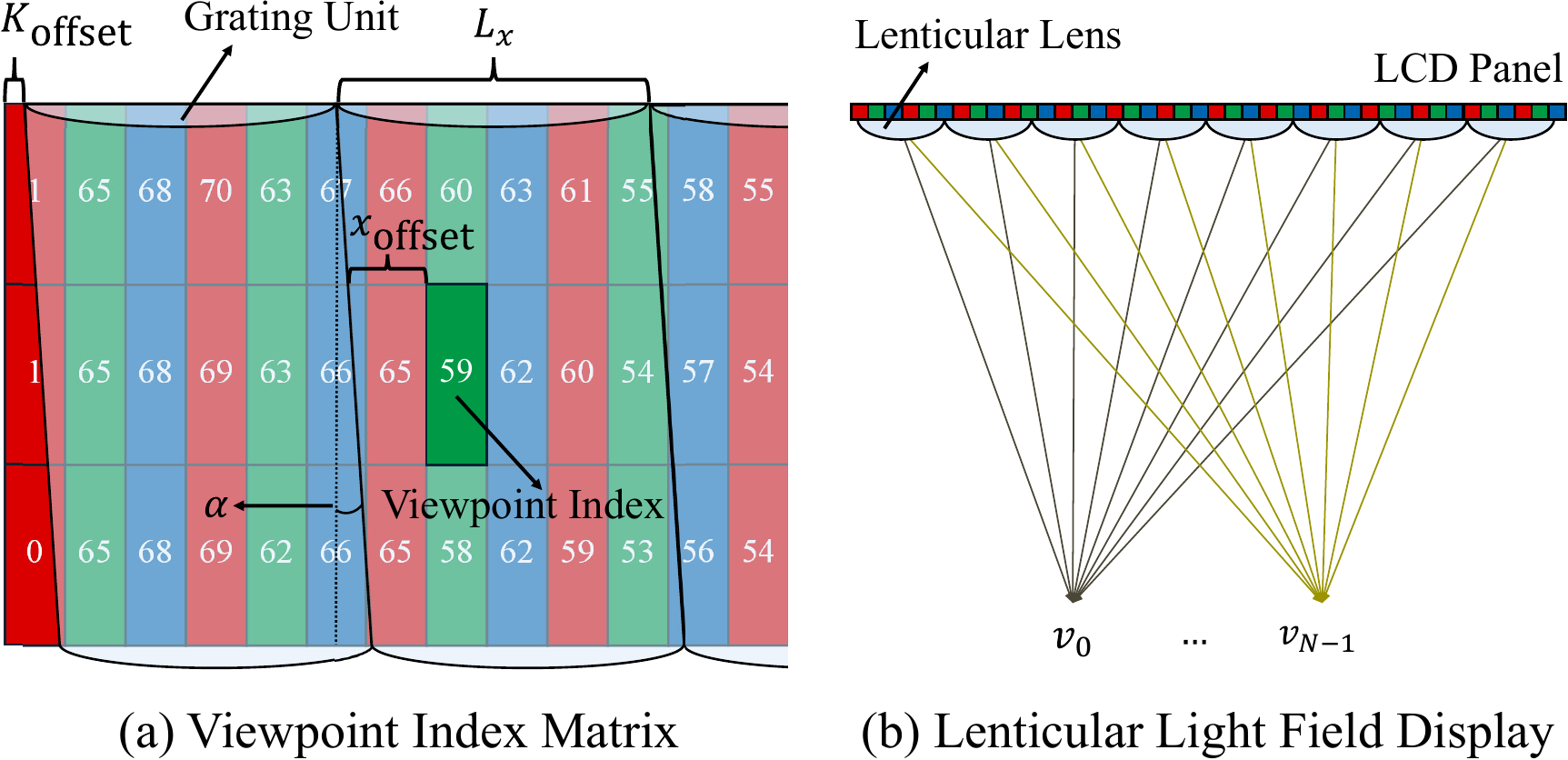}
    \caption{\textbf{Principle of Lenticular Light Field Displays.} (a) Viewpoint index matrix. Based on display parameters, each subpixel is assigned to a unique viewpoint index, forming the viewpoint index matrix $\mathbf{V}$. (b) Lenticular Light Field Display. The lenticular lens array refracts light from the LCD panel, directing each subpixel to a specific viewing angle.}
    \Description{light field display}
    \label{fig:lenticular_lfd_viewpoint_number_matrix}
\end{figure}
\section{Preliminaries}
\subsection{Light Field Display}


Our method operates directly on the interlaced subpixel layout of LFDs. To clarify this structure, we briefly review lenticular LFDs, which is the most widely adopted commercial realization of interlaced multi-view displays and serves as the running example throughout this paper.


Lenticular LFDs place a vertically oriented lenticular lens array above an LCD panel, as shown in \cref{fig:lenticular_lfd_viewpoint_number_matrix}(a). Each cylindrical lens spans multiple subpixels and refracts their emitted light into distinct directions, so that a viewer at a given position receives only the subpixels intended for that viewpoint (see \cref{fig:lenticular_lfd_viewpoint_number_matrix}(b)). This enables glasses-free 3D presentation via an interlaced image at panel resolution, where each subpixel encodes radiance from its designated viewpoint.


Constructing this interlaced image requires a viewpoint index matrix specifying, 
for every subpixel, which viewpoint it should represent (the numerical labels in 
\cref{fig:lenticular_lfd_viewpoint_number_matrix}(a)). Following the standard 
lenticular model, the index is determined by a subpixel's horizontal offset within 
a grating unit, governed by three display parameters: grating tilt angle $\alpha$, 
grating line count $L_x$ (in subpixel units), and lens-to-panel misalignment offset 
$K_{\mathrm{offset}}$.


For an LCD panel of width $W$ and height $H$, let $(x, y, u)$ denote a subpixel 
at column $x \in \{0,\dots,W-1\}$, row $y \in \{0,\dots,H-1\}$, and RGB channel 
$u \in \{0,1,2\}$. Its horizontal offset within the grating unit is:
\begin{align}
    d_{\mathrm{offset}} &= 3x + u + 3y\tan(\alpha) - K_{\mathrm{offset}}, \\
    x_{\mathrm{offset}} &= d_{\mathrm{offset}} \bmod L_x,
\end{align}
where the factor $3$ reflects the RGB subpixel layout. The viewpoint index is then:
\begin{align}
    j = \left\lfloor N \cdot \frac{x_{\mathrm{offset}}}{L_x} \right\rfloor,
\end{align}
with $N$ the total number of viewpoints. Collecting $j$ over all $(x,y,u)$ yields 
the viewpoint index matrix $\mathbf{V} \in \mathbb{Z}^{W \times H \times 3}$, 
uniquely determined by the display parameters.

\subsection{GPU Warps and Memory Coalescing}
\label{sec:prelim_gpu}

Modern Graphics Processing Units (GPUs) employ a Single Instruction, Multiple Threads (SIMT) architecture to achieve massive parallelism. In this execution model, individual threads are grouped into bundles known as \textit{warps} (typically comprising 32 threads in NVIDIA architectures). Threads within a warp execute the same instruction in a lock-step manner. While this architecture allows for high computational throughput, it imposes strict requirements on memory access patterns to maintain efficiency.

When a warp issues a memory request, the GPU memory controller services the warp's requests via \textit{memory transactions}, fetching data in aligned segments. To maximize throughput, the memory controller performs \textit{memory coalescing}, a process that consolidates memory requests from a warp into the minimum number of transactions.
This optimization takes effect only when the memory addresses accessed by a warp's threads are contiguous and aligned (\textit{coalesced access}). Conversely, scattered access patterns cause \textit{memory divergence}, forcing the controller to issue multiple separate transactions. This fragmentation significantly increases latency and wastes memory bandwidth, resulting in suboptimal hardware utilization.

\subsection{3D Gaussian Splatting}
\label{sec:prelim_3dgs}

We briefly review the 3DGS representation and its standard rendering pipeline.
In 3DGS, a scene is modeled as a set of anisotropic 3D Gaussians $\mathcal{G}=\{\mathcal{G}_i\}_{i=0}^{M-1}$, 
where each Gaussian $\mathcal{G}_i$ is defined by a 3D mean $\boldsymbol{\mu}_i$, 
a covariance matrix $\boldsymbol{\Sigma}_i$, an opacity $o_i$, and spherical harmonics (SH) coefficients $\mathbf{h}_i$.
The standard 3DGS renderer employs a tile‑based rasterization pipeline, where the image plane is partitioned into fixed‑size rectangular tiles (e.g., 16×16 pixels) for parallel processing. Rendering a view proceeds in a sequence of stages: (1) projection, (2) key generation, (3) sorting, and (4) alpha blending. Each Gaussian is projected onto the image plane, and its overlapping tiles are determined. A key is assigned to each Gaussian–tile pair, and the pair with its key is then sorted in depth order to produce a front‑to‑back sequence of splats for each tile, referred to as the Gaussian list. The key consists of tile ID and view‑space depth. The depth‑sorted splats are composited through alpha blending to yield the final pixel values, with each pixel’s blending executed independently within a single thread. The thread‑to‑pixel mapping inherently satisfies memory coalescing, ensuring efficient GPU execution.
While efficient for single‑view rendering, extending 3DGS to multi‑view rendering significantly increases computational cost, as the number of Gaussian–tile pairs grows proportionally with the number of viewpoints.

\section{\ours{}}
\label{sec:method}

Given a set of 3D Gaussians
$\mathcal{G}=\{\mathcal{G}_i\}_{i=0}^{M-1}$ and a 3D display setup,
\ours{} synthesizes a high-resolution interlaced light-field image
$\mathbf{I}_{LF} \in \mathbb{R}^{W \times H \times 3}$. The display setup
specifies the target viewpoints
$\mathcal{V}=\{v_i\}_{i=0}^{N-1}$ and the viewpoint index matrix
$\mathbf{V} \in \mathbb{Z}^{W\times H \times 3}$, which determines the
subpixel-to-viewpoint assignment for the interlaced panel.

\begin{figure*}[t]
    \centering 
    \includegraphics[width=0.9\textwidth]{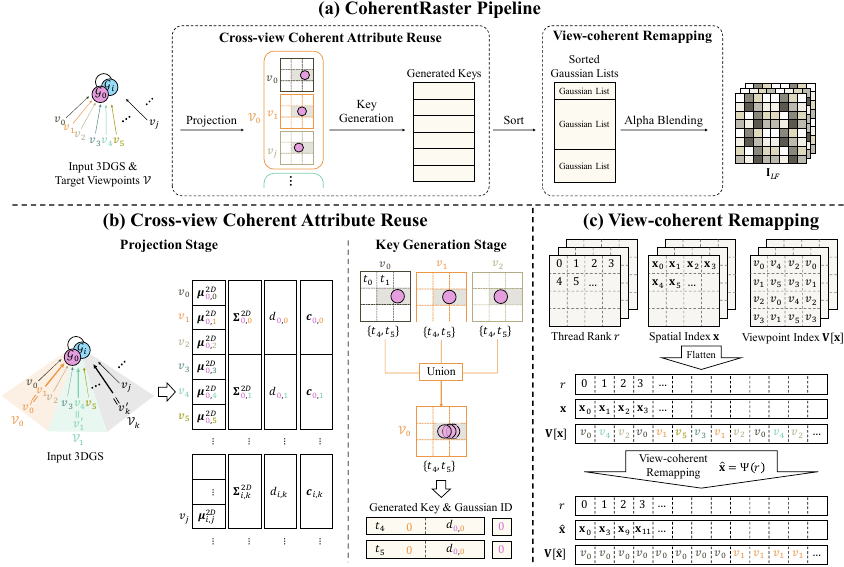}
    \caption{\textbf{Overall pipeline of proposed \ours{}.} The framework synthesizes high-resolution light field images from the input 3DGS and target viewpoints. In the projection and key generation stage, \viewreuse{} eliminates redundant computations by reusing projected attributes and generating sorting keys per cluster. Subsequently, during alpha blending, \viewremapping{} reorganizes thread execution based on viewpoint indices, thereby restoring coalesced memory access for efficient rendering.}
    \Description{pipeline}
    \label{fig:pipeline}
\end{figure*}

As illustrated in \cref{fig:pipeline}(a), \ours{} adapts the 3D Gaussian representation to the subpixel-level layout of light-field displays.  Unlike traditional light field coding methods, which render a full RGB image for each viewpoint and then sample subpixel values to construct an interlaced image, our approach directly determines, for every subpixel $(x,y,u)$, which Gaussians contribute to it and with what color. This eliminates the need to generate complete per‑view images, extending the rasterization process to operate at subpixel granularity and to aggregate contributions across multiple viewpoints.

\subsection{Subpixel-Level Rasterization}
\label{sec:subpixel_rasterization}

Light‑field displays interlace multiple viewpoints at the subpixel level, assigning each subpixel $(x,y,u)$ to a viewpoint index $\mathbf{V}[x,y,u]$. Consequently, rasterization requires computing Gaussian contributions not for a single view but simultaneously for multiple views, and not per pixel but per subpixel. To achieve this, we extend the tile-based rasterization process to support multi-view rendering at the subpixel level. We describe this extension across four stages: projection, key generation, sorting, and alpha blending.

\paragraph{Projection}
For each viewpoint $v_j$, every Gaussian $\mathcal{G}_i$ is projected onto the image plane, yielding its 2D mean $\boldsymbol{\mu}^{\text{2D}}_{i,j}$, covariance $\Sigma^{\text{2D}}_{i,j}$, depth $d_{i,j}$, and view-dependent color $\mathbf{c}_{i,j}$ as:
\begin{equation}
    \boldsymbol{\mu}^{\text{2D}}_{i,j}, \Sigma^{\text{2D}}_{i,j}, d_{i,j}, \mathbf{c}_{i,j} = \Pi(v_j;\mathcal{G}_i),
\end{equation}
where $\Pi(\cdot)$ denotes the projection operator. This produces $N$ sets of screen-space attributes per Gaussian, one for each viewpoint.

\paragraph{Key Generation}

In this stage, the overlapping tiles of each projected Gaussian are determined for every viewpoint $v_j$. A key is then assigned to each Gaussian-tile pair, augmented with the viewpoint index $j$. Thus, each key consists of the tile ID, the viewpoint ID $j$, and the view‑space depth $d_{i,j}$, providing the ordering information required for subsequent sorting.

\paragraph{Sorting}

The collected Gaussian-tile pairs are sorted in ascending order of view‑space depth, independently for each tile and viewpoint. This produces a front‑to‑back sequence of splats per tile for every viewpoint, forming a per‑tile, per‑view Gaussian list. These lists are stored in contiguous memory as a single unified Gaussian list, ordered by tile ID and viewpoint ID. Mapping each subpixel to its Gaussian list ensures that view-related Gaussian contributions are composited in the correct depth order within each tile.

\paragraph{Alpha Blending}
After depth sorting, the rasterizer traverses subpixels within each tile in row‑major order.  For each subpixel $(x,y,u)$, the splat sequence matching its tile index and viewpoint index $\mathbf{V}[x,y,u]$ is loaded. Gaussian contributions from this sequence are accumulated in front‑to‑back order through alpha blending, thereby producing the final light-field image $\mathbf{I}_{LF}$. 


Although conceptually straightforward, this naive subpixel-level pipeline exhibits two inefficiencies that severely limit throughput: redundant per-view evaluation and uncoalesced memory access under the interlaced subpixel layout. We address these bottlenecks with two complementary strategies: \viewreuse{} (\cref{subsec:viewreuse}) eliminates redundant computation across neighboring viewpoints in the projection, key generation, and sorting stages, while \viewremapping{} (\cref{subsec:viewremapping}) restores coalesced memory access during the alpha blending stage by reordering thread-to-subpixel mapping.

\subsection{\viewreuse{}}
\label{subsec:viewreuse}


Cross-view Coherent Attribute Reuse eliminates redundant per-view computation by reusing projected Gaussian attributes that vary smoothly across spatially adjacent viewpoints.
To achieve this, we group neighboring viewpoints into clusters and reuse their projected attributes within each cluster, as illustrated in \cref{fig:pipeline}(b). This clustering strategy is consistently applied across the projection, key generation, and sorting stages, significantly reducing per-view overhead.

Formally, we uniformly partition the full set of viewpoints $\mathcal{V}=\{v_i\}_{i=0}^{N-1}$ into $K$ disjoint clusters $\{\mathcal{V}_0,\dots,\mathcal{V}_{K-1}\}$, where $K < N$, and represent each cluster $\mathcal{V}_k$ by its geometric center view $v'_{k} \in \mathcal{V}_k$.
This clustering strategy mitigates the overhead of per-view evaluation that the subpixel-level 3DGS rasterization would otherwise incur. Further details on the partitioning and representative view selection, along with pseudocode for the full pipeline, are provided in the supplementary material.

\paragraph{Projection}
The 2D mean of a Gaussian changes noticeably with viewpoint because it corresponds to the projected center of the
3D ellipsoid. Even small viewpoint shifts cause the projected center to move across the image plane, which directly
affects tile assignment. To avoid geometric artifacts such as incorrect tile coverage, we therefore compute the 2D
mean of each Gaussian $\mathcal{G}_i$ independently for every view $v_j$:
\begin{equation}
    \boldsymbol{\mu}^{\text{2D}}_{i,j} = \Pi_{\mathrm{mean}}(v_j; \mathcal{G}_i).
\end{equation}

In contrast, other projected attributes—such as the 2D covariance, depth, and SH‑based color—vary much more
smoothly across nearby viewpoints. These quantities depend primarily on local surface orientation and shading,
which change gradually under small view shifts. Leveraging this cross‑view smoothness, we compute these attributes
once at the cluster representative view $v'_k$ and reuse them for all views $v_j \in \mathcal{V}_k \setminus v'_k$:
\begin{equation}
    \boldsymbol{\Sigma}^{\text{2D}}_{i,k} = \Pi_{\mathrm{cov}}(v'_k;\mathcal{G}_i), \quad
    d_{i,k} = \Pi_{\mathrm{depth}}(v'_k;\mathcal{G}_i), \quad
    \mathbf{c}_{i,k} = \Pi_{\mathrm{SH}}(v'_k;\mathcal{G}_i),
\end{equation}
where $\boldsymbol{\Sigma}^{\text{2D}}_{i,k}$, $d_{i,k}$, and $\mathbf{c}_{i,k}$ denote the 2D covariance, depth,
and SH‑based color of the $i$‑th Gaussian $\mathcal{G}_i$ observed from the $k$‑th cluster view $v'_k$, respectively,
and $\Pi_{\mathrm{cov}}(\cdot)$, $\Pi_{\mathrm{depth}}(\cdot)$, and $\Pi_{\mathrm{SH}}(\cdot)$ are the corresponding projection operators.

During the subsequent stages, the rasterizer accesses the per-view 2D mean via the viewpoint ID and retrieves the other shared attributes via the corresponding cluster ID. Using this combination, the contribution of each Gaussian is accurately evaluated and blended, ensuring both efficiency and high-fidelity rendering.

\paragraph{Key Generation} \label{para:key_generation}
\cref{fig:pipeline}(b) illustrates our key generation scheme. To begin, for each Gaussian, we include tiles that overlap with at least one viewpoint in the cluster. This produces per-cluster Gaussian-tile pairs rather than redundant per‑view ones, thereby reducing the total number of pairs to be subsequently processed.

Subsequently, we assign a single 64‑bit sorting key to each Gaussian-tile pair within every cluster. For each pair of Gaussian $\mathcal{G}_i$ and $t$-th tile, the key is constructed by packing the tile ID $t$, cluster ID $k$, and depth $d_{i,k}$ into a tuple as:
\begin{equation}
    \text{key}_{i,t} = (t, k, d_{i,k}),
\end{equation}
where $d_{i,k}$ denotes the depth of Gaussian $\mathcal{G}_i$ projected from the cluster center view $v'_k$. As the number of Gaussian-tile pairs with its keys is substantially reduced compared to per‑view key generation, the overall sorting workload decreases considerably, yielding significant computational savings.

\paragraph{Sorting}
Sorting the keys reorganizes the Gaussians into contiguous lists, each corresponding to a unique tile-cluster pair $(t, k)$, where $t$ and $k$ denote the tile and cluster IDs, respectively. Within each list, Gaussians are inherently ordered by depth due to the bitwise structure of the key. This layout allows the rasterizer to access the target list through the memory range $[S_{t, k}, E_{t, k})$, with $S_{t, k}$ and $E_{t, k}$ denoting the starting and ending offsets. The contiguous lists enable efficient front‑to‑back traversal of Gaussians during alpha blending.

\paragraph{Discussion}
While utilizing the per-cluster Gaussian-tile pair effectively reduces the key count, it can occasionally assign a Gaussian to a tile in which it is not actually visible
for some views within the cluster.
However, we find that the performance gain from reducing the sorting overhead significantly outweighs this cost, as analyzed in \cref{sec:ablation}. Furthermore, during the final alpha blending, the contribution of each Gaussian is precisely evaluated based on its mean and covariance and the rendered subpixel coordinate. Consequently, any unnecessary Gaussians yield negligible opacity and are naturally handled, preserving high-fidelity rendering quality.

\begin{table*}[t]
    \caption{\textbf{Quantitative Evaluation}. We evaluate rendering speed (FPS) and image quality (PSNR, SSIM, LPIPS) on the Synthetic Blender~\cite{nerf} and Mip-NeRF 360~\cite{mipnerf360} datasets. $|\mathcal{V}_k|=16$ for the 63-view 2K setup and $|\mathcal{V}_k|=18$ for the 71-view 4K setup are selected to ensure balanced view partitioning for each display specification.}
    
    \label{table:quantitative_evaluation}
    \resizebox{0.9\textwidth}{!}{
        \begin{tabular}{l|cccc|cccc|cccc|cccc}
        \Xhline{2\arrayrulewidth}
        \makecell[c]{\multirow{3}{*}{Method}} & \multicolumn{8}{c|}{Synthetic Blender}                                                            & \multicolumn{8}{c}{Mip-NeRF 360}                                                             \\ 
        \cline{2-17}
        
         & \multicolumn{4}{c|}{2K ($1440 \times 2560$)}                                   & \multicolumn{4}{c|}{4K ($3840 \times 2160$)}              & \multicolumn{4}{c|}{2K ($1440 \times 2560$)}                                 & \multicolumn{4}{c}{4K ($3840 \times 2160$)}              \\
                   & FPS$\uparrow$  & PSNR$\uparrow$  & SSIM$\uparrow$   & \multicolumn{1}{c|}{LPIPS$\downarrow$}  & FPS$\uparrow$  & PSNR$\uparrow$  & SSIM$\uparrow$   & LPIPS$\downarrow$  & FPS$\uparrow$  & PSNR$\uparrow$  & SSIM$\uparrow$  & \multicolumn{1}{c|}{LPIPS$\downarrow$} & FPS$\uparrow$  & PSNR$\uparrow$  & SSIM$\uparrow$  & LPIPS$\downarrow$ \\ 
        \hline
        \hline
        3DGS       & 5.8  &  \multicolumn{3}{c|}{\text{pseudo ground-truth}}      & 4.1  & \multicolumn{3}{c|}{\text{pseudo ground-truth}}      & 3.9  & \multicolumn{3}{c|}{\text{pseudo ground-truth}}     & 2.1  & \multicolumn{3}{c}{\text{pseudo ground-truth}} \\ 
        Ours ($|V_k|=2$)  & 54 & 62.13 & 0.9997 & 0.0003 & 36 & 62.08 & 0.9998 & 0.0004 & 20 & 53.08 & 0.998 & 0.001 & 11 & 53.35 & 0.998 & 0.001 \\
        Ours ($|V_k|=4$)  & 75 & 56.75 & 0.9994 & 0.0005 & 49 & 56.86 & 0.9995 & 0.0005 & 28 & 47.44 & 0.995 & 0.003 & 15 & 48.27 & 0.996 & 0.002 \\
        Ours ($|V_k|=8$)  & 88 & 51.94 & 0.9989 & 0.0009 & 56 & 52.19 & 0.9991 & 0.0010 & 30 & 42.74 & 0.990 & 0.008 & 16 & 43.78 & 0.992 & 0.006 \\
        Ours ($|V_k|=16$) & 84 & 46.86 & 0.9976 & 0.0024 & -    & -     & -      & -      & 24 & 37.50 & 0.974 & 0.025 & -    & -     & -     & -     \\
        Ours ($|V_k|=18$) & -    & -     & -      & -      & 52 & 46.40 & 0.9976 & 0.0027 & -    & -     & -     & -     & 13 & 38.00 & 0.977 & 0.021 \\ 
        \Xhline{2\arrayrulewidth}
        \end{tabular}
        
    }

\end{table*}

\subsection{\viewremapping{}}
\label{subsec:viewremapping}
\viewremapping{} resolves the uncoalesced memory access in subpixel‑level rasterization by reordering how GPU threads are assigned to subpixels. Instead of mapping thread ranks to spatial indices in raster order, \viewremapping{} sorts the subpixel coordinates by their viewpoint indices and assigns them to threads accordingly. This ensures that threads within a warp process subpixels belonging to the same or nearby viewpoints, restoring viewpoint monotonicity and enabling coalesced memory access.

Conventional tile‑based rasterization pipelines assume that spatially adjacent subpixels share similar data requirements~\cite{pulsar,3dgs}. In light‑field displays, however, the lens geometry interleaves viewpoints across the panel (see \cref{fig:lenticular_lfd_viewpoint_number_matrix}). As a result, even threads mapped to the same tile may diverge in viewpoint index and must access different Gaussian lists organized by tile and cluster (\cref{para:key_generation}). Grouping subpixels by viewpoint before assigning them to threads ensures that warp threads access the same or nearby Gaussian list, significantly reducing bandwidth overhead.

\cref{fig:pipeline}(c) illustrates the construction of the viewpoint‑sorted mapping. For each tile, we first linearize the subpixel coordinates $\mathbf{x}$ in row‑major order and sort them by their viewpoint indices $\mathbf{V}[\mathbf{x}]$. The reordered indices are stored in a lookup table $\Psi$, which is precomputed once since the lens geometry is fixed. During rasterization, each thread accesses subpixels through this table, which enforces viewpoint monotonicity within the warp. For any two consecutive threads with ranks $r$ and $r+1$, the viewpoint indices satisfy:
\begin{equation}
    \mathbf{V}[\Psi(r)] \le \mathbf{V}[\Psi(r+1)].
\end{equation}
This ordering increases the likelihood that neighboring threads process subpixels associated with the same or adjacent viewpoints, which implies identical or adjacent cluster IDs. Since Gaussian lists are organized by tile ID and then by cluster ID, threads naturally access the same or nearby Gaussian list, restoring coalesced memory access.

With the mapping $\Psi$, the rasterization kernel performs alpha blending with optimized memory access. A thread with rank $r$ retrieves its spatial index $\hat{\mathbf{x}} = (x, y, u) = \Psi(r)$ and the corresponding viewpoint index $j = \mathbf{V}[\hat{\mathbf{x}}]$. It then identifies the Gaussian list defined by the memory range $[S_{t,k}, E_{t,k})$, where $t$ and $k$ denote the tile and cluster IDs associated with $\hat{\mathbf{x}}$ and viewpoint index $j$. For each Gaussian $\mathcal{G}_i$ in this list, the thread evaluates its contribution using the projected attributes from \cref{subsec:viewreuse}, namely the 2D covariance $\boldsymbol{\Sigma}^{\text{2D}}_{i,k}$ and the color vector $\mathbf{c}_{i,k}$. The final subpixel intensity is accumulated as:
\begin{equation}
    C(\hat{\mathbf{x}}) = \sum_{i \in \mathcal{N}} c^{(u)}_{i,k} \alpha_i \prod_{p=1}^{i-1} (1 - \alpha_p),
\end{equation}
\begin{equation}
    \alpha_i = o_i \cdot \exp\left(-\frac{1}{2}(\hat{\mathbf{x}} - \boldsymbol{\mu}^{\text{2D}}_{i,j})^\intercal
    {\boldsymbol{\Sigma}^{\text{2D}}}^{-1}_{i,k}(\hat{\mathbf{x}} - \boldsymbol{\mu}^{\text{2D}}_{i,j})\right),
\end{equation}
where $\mathcal{N}$ is the ordered set of Gaussians in the list, $c^{(u)}_{i,k}$ is the color component for channel $u$, and $o_i$ is the learned opacity. The viewpoint index $j$ is used for the mean $\boldsymbol{\mu}^{\text{2D}}_{i,j}$, while the cluster ID $k$ is used for the reused covariance $\boldsymbol{\Sigma}^{\text{2D}}_{i,k}$ and color $\mathbf{c}_{i,k}$. The process terminates when the accumulated opacity saturates or the list is exhausted. The resulting intensity $C(\hat{\mathbf{x}})$ is written directly to the light‑field image $\mathbf{I}_{LF}$ at the corresponding subpixel location, producing the interlaced output without additional post‑processing.

\section{Experiments}
\label{sec:experiments}

\subsection{Experimental Setup}

\paragraph{Implementation Details}

\ours{} is built upon \texttt{gsplat}~\cite{gsplat}, an open-source library for Gaussian splatting, with customized CUDA kernels tailored to our rendering pipeline. In addition, we integrate the AccuTile~\cite{speedy_splat} algorithm into both our framework and the baselines by default to ensure accurate Gaussian–tile intersection. For 3DGS optimization, we follow the standard training configurations of \texttt{gsplat}, with the regularization scheme of 3DGS-MCMC~\cite{kheradmand2024mcmc} applied to real-world scenes. We use cluster size $|\mathcal{V}_k|=8$ as default setting for rendering. All experiments are conducted on an NVIDIA RTX 5090 GPU (32GB).

\paragraph{Datasets and Evaluation Protocol}
To evaluate the performance of \ours{}, we use eight synthetic scenes from the Synthetic Blender dataset~\cite{nerf} and seven real-world scenes from the Mip-NeRF 360 dataset~\cite{mipnerf360}. 


Since interlaced images for LFDs require dense adjacent views, we synthesize four test trajectories per scene by orbiting the center, with cameras consistently oriented inward to cover a diverse range of viewing angles. Evaluations are conducted under two display configurations: (1) a landscape setup, requiring 71 views within a $53^\circ$ range at 4K ($3840 \times 2160$), and (2) a portrait setup, requiring 63 views within a $53^\circ$ range at 2K ($1440 \times 2560$). 
We use a Looking Glass Go for 2K rendering and a Looking Glass 16" Light Field Display for 4K rendering.
To align the dataset with these display specifications, we adjust camera intrinsics: expanding the FOV for Synthetic Blender to include peripheral regions that are originally out of frame, and narrowing the FOV for Mip-NeRF 360 to fit the target resolutions, cropping the original image coverage.

To evaluate image quality of \ours{}, we measure PSNR, SSIM, and LPIPS~\cite{lpips} on per-view images.
Since ground-truth images for the test trajectories are unavailable, we utilize images rendered by original 3DGS as pseudo ground-truth to assess fidelity.
For rendering speed evaluation, we report the average frames per second (FPS) for generating the final interlaced images along each test trajectory.

\paragraph{Baselines}

We compare \ours{} against prior light field rendering approaches, all re‑implemented in the \texttt{gsplat}~\cite{gsplat} framework. For 3DGS~\cite{3dgs} and its AccuTile~\cite{speedy_splat} variant, we adopt a full‑frame rendering pipeline that renders all multi‑view images at full resolution before interlacing. Our comparison further includes two representative categories of prior work: subpixel‑based and MPI‑based approaches. Since existing LFD rendering methods such as \citet{text_driven_lf_editing_gs} and \citet{g2lf} do not provide code, we reproduce the core ideas underlying each approach rather than attempting an exact re‑implementation.

For the subpixel‑based baseline, which also forms the foundation of our method, we adopt the subpixel sampling scheme described in \cref{sec:subpixel_rasterization} without cross‑view coherent attribute reuse or view‑coherent remapping. We refer to this variant as Subpixel‑3DGS. For the MPI‑based baseline, we construct 64 MPI planes from a reference camera covering all views, computed at a $4\times$ downscaled resolution to reduce memory overhead. During interlaced rendering, we similarly apply subpixel‑level sampling by selecting the required subpixels from the planes and aggregating them via alpha blending.






\begin{table}[t]
    \caption{\textbf{Comparison with Baselines}. \ours{} outperforms the baselines under both 2K and 4K display configurations.}
    \label{table:comparison_with_baselines}
    
    \resizebox{1.0\linewidth}{!}{

\begin{tabular}{c|c|cc|cc}
\Xhline{2\arrayrulewidth}
\multirow{2}{*}{Rendering}                       &   \multirow{2}{*}{Method}  & \multicolumn{2}{c|}{Synthetic Blender} & \multicolumn{2}{c}{Mip-NeRF 360} \\
\cline{3-6}
                                                 &                      & \multicolumn{1}{c|}{2K FPS}      & 4K FPS     & \multicolumn{1}{c|}{2K FPS}   & 4K FPS   \\
\hline \hline
\multicolumn{1}{c|}{\multirow{2}{*}{Full-Frame}} & 3DGS           & \multicolumn{1}{c|}{5.8}     & 4.1     & \multicolumn{1}{c|}{3.9}  & 2.1  \\
\multicolumn{1}{c|}{}                            & 3DGS (batch=36) & \multicolumn{1}{c|}{20}    & 13    & \multicolumn{1}{c|}{7.4}  & 4.0  \\ \hline
\multicolumn{1}{c|}{\multirow{3}{*}{Subpixel}}   & Subpixel-3DGS   & \multicolumn{1}{c|}{28}    & 19    & \multicolumn{1}{c|}{11} & 5.7  \\
\multicolumn{1}{c|}{}                            & MPI             & \multicolumn{1}{c|}{0.8 }     & 0.4      & \multicolumn{1}{c|}{0.8 }  & 0.4   \\
\multicolumn{1}{c|}{}                            & Ours            & \multicolumn{1}{c|}{88 }    & 56     & \multicolumn{1}{c|}{30  } & 16  \\

\Xhline{2\arrayrulewidth}
\end{tabular}

    }
    
\end{table}

\subsection{Results and Analysis}

\paragraph{Evaluation}

\cref{table:quantitative_evaluation} summarizes the quantitative evaluation of our framework. \ours{} achieves real-time rendering speed of high-resolution light field rendering, enabling interactive 3D visualization. Compared to the original 3DGS, \ours{} demonstrates faithful rendering results with only negligible quality degradation, despite substantial improvement in rendering efficiency. The number of viewpoints in each cluster determines the trade-off between efficiency and quality: as the count increases, rendering speed improves, but high-fidelity rendering cannot be fully preserved. $|\mathcal{V}_k|=8$ achieves the best performance across both datasets.

For the Synthetic Blender dataset, we demonstrate real-time rendering speed for both 2K and 4K LFDs, whereas the original 3DGS rasterization pipeline suffers from redundant multi-view computation. Likewise, our framework achieves real-time rendering at 2K resolution and feasible rendering at 4K resolution on the Mip-NeRF 360 dataset. Notably, our method is the first to enable real-time rendering beyond 2K resolution for real-world 3D scenes, underscoring its practical significance.

\paragraph{Comparison}


\cref{table:comparison_with_baselines} presents a quantitative comparison of rendering speed against baseline methods, demonstrating that \ours{} achieves superior performance over existing light field rendering approaches. 
Full-frame pipelines, including 3DGS and its batched variant, exhibit limited efficiency due to wasteful computation on non-contributing pixels; although batching viewpoints improves throughput, real-time rendering remains unattainable. 
Subpixel-3DGS improves efficiency by rendering only the necessary subpixels. Despite this improvement, it remains sub-optimal due to redundant computations across adjacent views and uncoalesced memory access patterns.
The MPI-based framework scales poorly with output resolution, as per-pixel traversal of the depth plane stack incurs rendering latency. Moreover, MPI method yields PSNR of 20.84 / 21.28 dB on Synthetic Blender and 19.48 / 19.11 dB on Mip-NeRF 360 at 2K / 4K, whereas CoherentRaster achieves higher PSNR across all cluster sizes (\cref{table:quantitative_evaluation}). This degradation stems from discretization artifacts caused by the limited number of depth planes. While increasing the plane count can suppress these artifacts, it further amplifies the latency.
In contrast, \ours{} achieves a speedup of $7.6\times$ compared to the full-frame 3DGS rendering pipeline while preserving high-fidelity rendering.

\begin{figure*}[t]
    \centering 
    \includegraphics[width=0.97\textwidth]{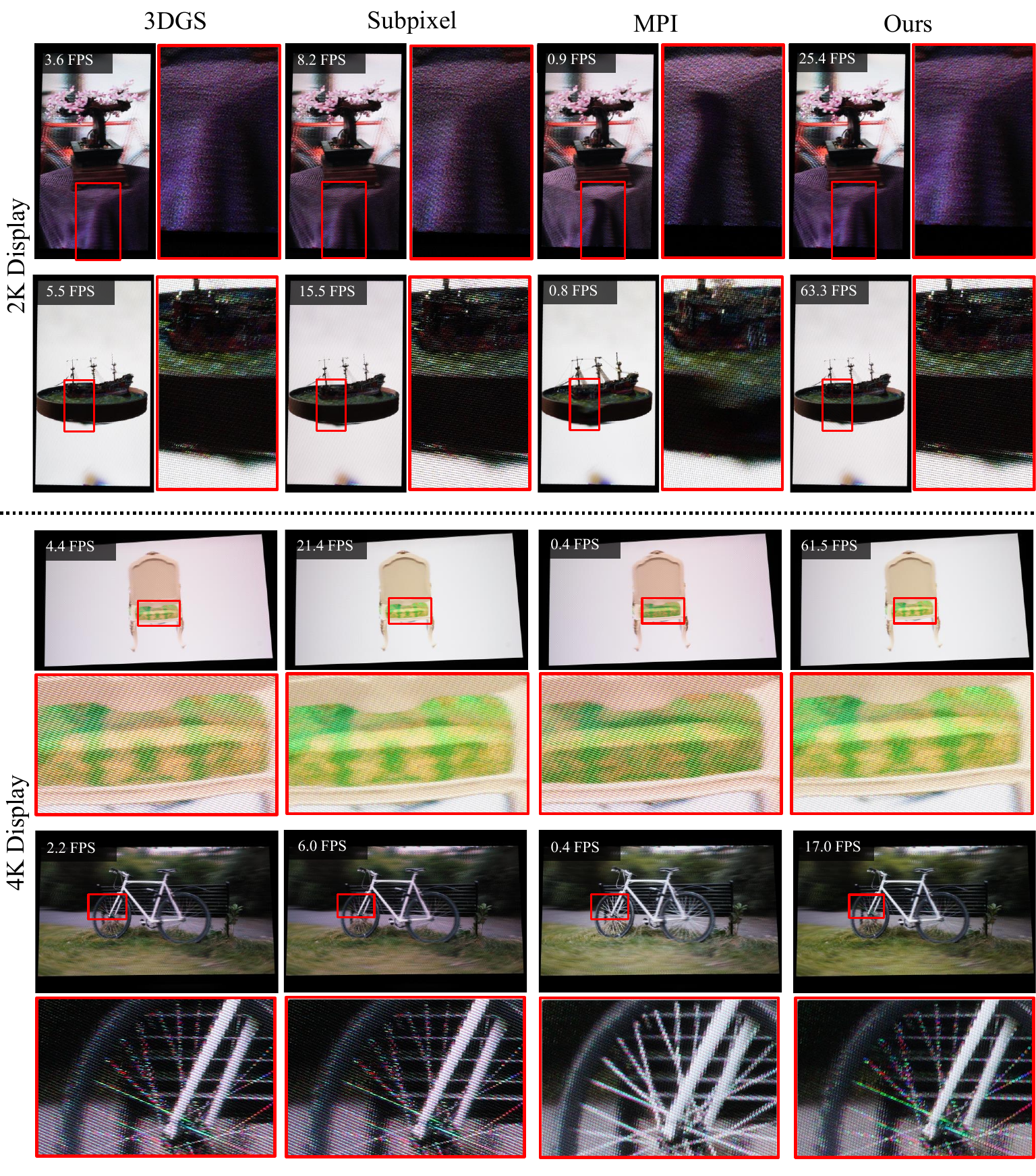}
    \caption{\textbf{Qualitative comparison on the light field display}. We present photographs captured directly from the display to compare the visual quality of \ours{} with the baseline. Our method achieves real-time frame rates while maintaining perceptual quality indistinguishable from the high-cost full-frame rendering. Note that slight color shifts or misalignments may appear due to the capture process.
    (Rows 1 and 4: \textit{Bonsai} and \textit{Bicycle} scenes from the Mip-NeRF 360 dataset; Rows 2 and 3: \textit{Ship} and \textit{Chair} scenes from the Synthetic Blender dataset)
    }
    \Description{rendering quality}
    \label{fig:display_quality}
\end{figure*}

\Cref{fig:display_quality} presents a qualitative comparison of the rendering results.
The MPI-based approach struggles to maintain visual fidelity; it exhibits discretization artifacts due to the limited number of depth planes, and suffers from geometric inconsistencies in peripheral views caused by warping from a single reference perspective.
In contrast, \ours{} achieves superior visual quality, generating sharp imagery that matches the full-frame baseline, even in complex regions.
Crucially, our method delivers this high fidelity while significantly outperforming both the full-frame and MPI baselines in terms of rendering speed.

\begin{table}[t]
    \caption{\textbf{Ablation study on the overall pipeline}. Our proposed strategies alleviate redundant computation and irregular memory access.}
    \label{table:ablation_study}
    
    \resizebox{1.0\linewidth}{!}{
    
    \begin{tabular}{l|c|c|c|c}
    \Xhline{2\arrayrulewidth}
    \makecell[c]{\multirow{2}{*}{Method}}                      & \multicolumn{2}{c|}{Synthetic Blender} & \multicolumn{2}{c}{Mip-NeRF 360}  \\
    \cline{2-5}
                                                 & 2K FPS     & 4K FPS      & 2K FPS   & 4K FPS    \\ 
    \hline
    \hline
    Ours w/o Reuse w/o Remap     & 28    & 19    & 11 & 5.7   \\
    Ours w/o Reuse               & 34    & 22    & 12 & 6.4   \\
    Ours w/o Remap               & 67    & 41    & 21 & 11 \\
    Ours                         & 88    & 56    & 30 & 16  \\
    \Xhline{2\arrayrulewidth}
    \end{tabular}
    }
    
\end{table}

\paragraph{Ablation Study}
\label{sec:ablation}
We validate the effectiveness of two key components of \ours{}: \viewreuse{} and \viewremapping{}. \Cref{table:ablation_study} reports the performance evaluation of each component. Without \viewreuse{}, substantial cross-view computations are required, preventing efficient multi-view rendering of complex 3D scenes. Without \viewremapping{}, inefficient memory coalescing hinders real-time rendering on real-world data. Each component individually contributes noticeable gains: \viewreuse{} improves throughput by reducing redundant computations, while \viewremapping{} further enhances memory access efficiency. When combined, their complementary effects enable real-time rendering even for high-resolution real-world scenes. For a detailed rendering time breakdown, see the supplemental document.


\section{Conclusion}
\label{sec:conclusion}

In this paper, we presented \ours{}, a real-time 3DGS-based light field rendering framework that leverages subpixel-level rasterization for high-resolution multi-view rendering. Our \viewreuse{} scheme eliminates the computational redundancy of dense view synthesis by reusing inter-view attributes, while \viewremapping{} resolves the uncoalesced memory access issue inherent to subpixel-level rasterization. By addressing these challenges, \ours{} achieves real-time performance on high-resolution LFDs with high fidelity.

\paragraph{Limitations.}
Despite achieving real-time performance, \ours{} relies on local consistency within view clusters. As a result, scenes with high-frequency specular effects may exhibit visual artifacts due to the \viewreuse{} strategy. Moreover, our current framework is restricted to static scenes; future work will focus on extending the pipeline to dynamic content.

\begin{acks}
We thank Jiyun Won for valuable assistance in the visualization of the experimental results. This research was supported by the Project of the Electronics and Telecommunications Research Institute (Development of the Core Technology for Artificial Intelligence-Based Light-field Image Generation and Quality Evaluation Technology for Light-field Display, 25YC1600); the National Research Foundation of Korea (NRF) grant funded by the Korea government (MSIT) (No. RS-2026-25492695); the Institute of Information \& Communications Technology Planning \& Evaluation (IITP) grant funded by the Korea government (MSIT) (No. RS-2026-25517417, Development of Compression, Reconstruction, and Rendering Technologies for Free-viewpoint Media); and the IITP grant funded by the Korea government (MSIT) (No. RS-2019-II191906, Artificial Intelligence Graduate School Program (POSTECH)). 
\end{acks}

\begin{figure*}[t]
    \centering 
    \includegraphics[width=1.0\textwidth]{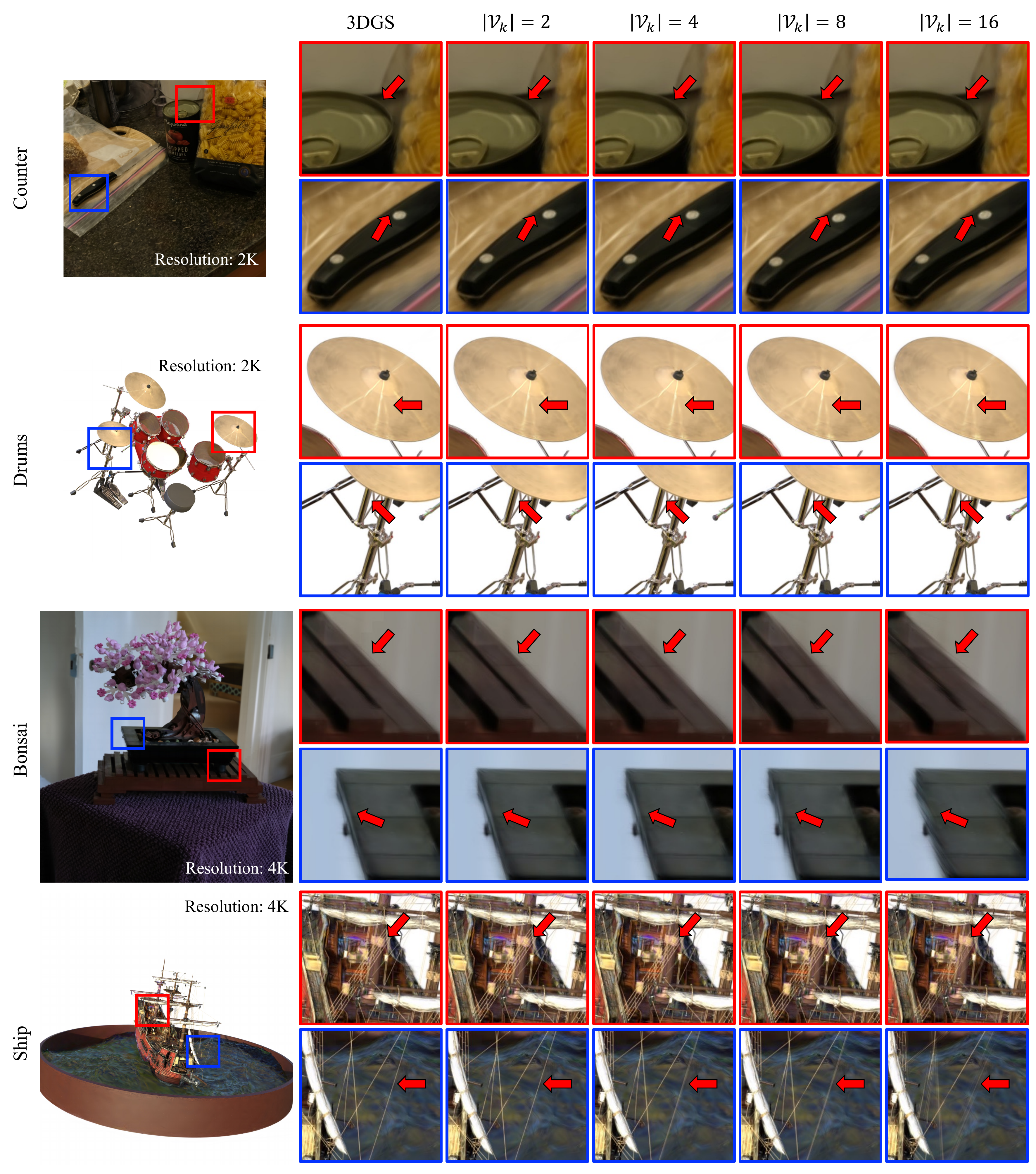}
    \caption{\textbf{Ablation study on cluster size}. Across different cluster sizes, our method maintains feasible visual quality without noticeable distortion.
    (\textit{Counter} and \textit{Bonsai} scenes from the Mip-NeRF 360 dataset; \textit{Drums} and \textit{Ship} scenes from the Synthetic Blender dataset)
    }
    \Description{rendering quality of each cluter size}
    \label{fig:cluster_quality}
\end{figure*}

\bibliographystyle{ACM-Reference-Format}
\bibliography{sections/reference}

\clearpage

\appendix

\section{Supplementary Material}
We provide a supplementary video demonstrating real-time rendering results with interactive camera control.

\section{Implementation Details}

\subsection{Viewpoint Clustering}
\label{sec:supp_cluster}

We uniformly partition the target viewpoints $\mathcal{V}=\{v_i\}_{i=0}^{N-1}$ into $K$ disjoint clusters. Specifically, the cameras are indexed sequentially along the viewing trajectory and grouped into contiguous clusters of size $|\mathcal{V}_k|$. In cases where $N$ is not perfectly divisible by the cluster size $|\mathcal{V}_k|$, the final cluster is padded by simply duplicating the last camera $v_{N-1}$. Within each cluster $\mathcal{V}_k$, we designate the representative view $v'_k$ as the median-index camera, corresponding to the camera located at local index $\lfloor |\mathcal{V}_k| / 2 \rfloor$ within the clustered sequence.

\subsection{Rasterization Algorithm}
\label{sec:alg_details}

Algorithms \ref{alg:ours_rasterization_main} and \ref{alg:ours_rasterization_sub} outline the detailed procedure of our proposed rasterization pipeline. Building upon the standard 3D Gaussian Splatting rasterizer, our pipeline efficiently synthesizes the final interlaced light field image $\mathbf{I}_{LF}$ in four stages:
projection and key generation with \viewreuse{}, sorting, and alpha blending with \viewremapping{}

\paragraph{Projection.}
We separate view-dependent attributes from cluster-shared ones. The 2D mean $\boldsymbol{\mu}^{\textnormal{2D}}_{i,j}$ is computed per view $v_j$, whereas the 2D covariance $\boldsymbol{\Sigma}^{\textnormal{2D}}_{i,k}$, depth $d_{i,k}$, and color $\boldsymbol{c}_{i,k}$ are computed only per cluster center $v'_k$ and reused across all views in the same cluster.

\begin{algorithm}[t]
    \caption{\ours{} rasterization pipeline (main)}
    \label{alg:ours_rasterization_main}
    \DontPrintSemicolon
    \SetKwInOut{Input}{Input}
    \SetKwInOut{Output}{Output}
    \SetKwProg{Fn}{Function}{:}{}
    \Input{
        $M$: \# Gaussians;\ $N$: \# views;\ $K$: \# clusters\\
        $\boldsymbol{\mu}, \boldsymbol{\Sigma}, o, \textnormal{SH}$: 3D means, covariances, opacities, SH coefficients\\
        $\mathcal{V}{=}\{v_j\}_{j=0}^{N-1}$: views;\ $\mathcal{V}'{=}\{v'_k\}_{k=0}^{K-1}$: cluster centers\\
        $\mathcal{T}$: tile IDs;\ $\mathbf{V}$: viewpoint index matrix
    }
    \Output{$\mathbf{I}_{LF}$: interlaced light-field image}
    \hrule
    \BlankLine
    \Fn{$\textnormal{Rasterization}(M, \boldsymbol{\mu}, \boldsymbol{\Sigma}, o, \textnormal{SH}, \mathcal{V}, \mathcal{V}', \mathcal{T})$}{
        \tcp{\textbf{Stage 1: Projection with Cross-view Coherent Attribute Reuse}}
        $\boldsymbol{\mu}^{\textnormal{2D}} \gets \emptyset$\;
        $\boldsymbol{\Sigma}^{\textnormal{2D}}, d, \boldsymbol{c} \gets \emptyset$\;
        \For(\tcp*[f]{in parallel over $M \times N$ threads}){$r \gets 0$ \KwTo $M \times N-1$}{
            $i, j \gets \textnormal{GetIndex}(r)$\;
            $\boldsymbol{\mu}^{\textnormal{2D}}_{i,j} \gets \Pi_{\textnormal{mean}}(v_j;\, \boldsymbol{\mu}_i)$\;
        }
        \For(\tcp*[f]{in parallel over $M \times K$ threads; reused across views}){$r \gets 0$ \KwTo $M \times K-1$}{
            $i, k \gets \textnormal{GetIndex}(r)$\;
            $\boldsymbol{\Sigma}^{\textnormal{2D}}_{i,k},\, d_{i,k},\, \boldsymbol{c}_{i,k} \gets \Pi_{\textnormal{cov, depth, SH}}(v'_k;\, \boldsymbol{\Sigma}_i, \textnormal{SH}_i)$\;
        }
        \BlankLine
        \tcp{\textbf{Stage 2: Key Generation with Cross-view Coherent Attribute Reuse}}
        $\mathcal{P} \gets \emptyset$\;
        \For(\tcp*[f]{in parallel over $M \times K$ threads}){$r \gets 0$ \KwTo $M \times K - 1$}{
            $\mathcal{K}_{i,k} \gets \textnormal{GenerateKeys}(\boldsymbol{\mu}^{\textnormal{2D}}, \boldsymbol{\Sigma}^{\textnormal{2D}}, d, \mathcal{T}, r)$\;
            $\mathcal{P} \gets \textnormal{SaveToBuffer}(\mathcal{K}_{i,k})$\;
        }
        \BlankLine
        \tcp{\textbf{Stage 3: Sort by (tile $t$, cluster $k$, depth $d$)}}
        $\mathcal{L} \gets \textnormal{Sort}(\mathcal{P})$\;
        \BlankLine
        \tcp{\textbf{Stage 4: Alpha Blending with View-coherent Remapping}}
        $\mathbf{I}_{LF} \gets \mathbf{0}$\;
        \For(\tcp*[f]{in parallel over pixel threads}){$r \gets 0$ \KwTo $W \times H \times 3 - 1$}{
            $\mathbf{I}_{LF} \gets \textnormal{Alpha-Blend}(\mathbf{I}_{LF}, \mathbf{V}, \mathcal{L}, \mathcal{T}, \boldsymbol{\mu}^{\textnormal{2D}}, \boldsymbol{\Sigma}^{\textnormal{2D}}, \boldsymbol{c}, o, r)$\;
        }
        \Return $\mathbf{I}_{LF}$
    }
\end{algorithm}

\paragraph{Key Generation.}
For each Gaussian, we compute its bounding box and identify overlapping tiles. A unique sorting key is generated in \texttt{GenerateKeys} with the following bit layout:
\begin{equation}
    \textnormal{key} = (t \ll (32+\textnormal{Bit}_K)) \mid (k \ll 32) \mid (\overline{d_{i,k}}),
\end{equation}
where $t$, $k$ represent the tile ID and Cluster ID, respectively. By placing the Tile ID at the most significant bits, followed by the Cluster ID and the bitwise integer representation of the depth $\overline{d_{i,k}}$, Gaussians are sorted primarily by spatial location, then by cluster, and finally by depth.

\paragraph{Sorting.}
All keys are saved into a global buffer $\mathcal{P}$ and sorted to produce the ordered list $\mathcal{L}$, grouping Gaussians by (tile, cluster, depth).

\begin{algorithm}[t]
    \caption{\ours{} rasterization pipeline (subroutines)}
    \label{alg:ours_rasterization_sub}
    \DontPrintSemicolon
    \SetKwProg{Fn}{Function}{:}{}
    \Fn{$\textnormal{GenerateKeys}(\boldsymbol{\mu}^{\textnormal{2D}}, \boldsymbol{\Sigma}^{\textnormal{2D}}, d, \mathcal{T}, r)$}{
        $i, k \gets \textnormal{GetIndex}(r)$\;
        $\mathcal{T}_{i,k} \gets \emptyset$ \tcp*[r]{Merge tiles over all views in cluster $\mathcal{V}_k$}
        \For{$l \gets 0$ \KwTo $|\mathcal{V}_k|-1$}{
            $j \gets \textnormal{GetViewpointIndex}(k, l)$\;
            $\textnormal{region} \gets \textnormal{ComputeRegionAccuTile}(\boldsymbol{\mu}^{\textnormal{2D}}_{i,j},\, \boldsymbol{\Sigma}^{\textnormal{2D}}_{i,k})$\;
            \ForEach{tile $t \in \mathcal{T}$ overlapped by $\textnormal{region}$}{
                $\mathcal{T}_{i,k} \gets \mathcal{T}_{i,k} \cup \{t\}$\;
            }
        }
        
        $\mathcal{K}_{i,k} \gets \emptyset$\;
        \ForEach{$t \in \mathcal{T}_{i,k}$}{
            $\textnormal{key} \gets (t \ll (32{+}\textnormal{Bit}_K)) \mid (k \ll 32) \mid \overline{d_{i,k}}$\;
            $\mathcal{K}_{i,k} \gets \textnormal{Append}(\mathcal{K}_{i,k},\, \langle \textnormal{key},\, i \rangle)$\;
        }
        \Return $\mathcal{K}_{i,k}$
    }
    \BlankLine
    \Fn{$\textnormal{Alpha-Blend}(\mathbf{I}_{LF}, \mathbf{V}, \mathcal{L}, \mathcal{T}, \boldsymbol{\mu}^{\textnormal{2D}}, \boldsymbol{\Sigma}^{\textnormal{2D}}, \boldsymbol{c}, o, r)$}{
        $\hat{\mathbf{x}} \gets \Psi_{\textnormal{remap}}(r)$ \tcp*[r]{View-coherent Remapping}
        $t \gets \textnormal{IdentifyTileID}(\hat{\mathbf{x}}, \mathcal{T})$\;
        $j \gets \mathbf{V}[\hat{\mathbf{x}}]$ \tcp*[r]{Per-subpixel viewpoint index}
        $k \gets \textnormal{GetClusterID}(j)$\;
        $[S_{t,k}, E_{t,k}) \gets \textnormal{IdentifyGaussianList}(\mathcal{L}, t, k)$\;
        $p \gets 0$\;
        \For{$e \gets S_{t,k}$ \KwTo $E_{t,k}-1$}{
            $i \gets \mathcal{L}[e]$\;
            \tcp{Per-view $\boldsymbol{\mu}^{\textnormal{2D}}_{i,j}$; reused $\boldsymbol{\Sigma}^{\textnormal{2D}}_{i,k}, \boldsymbol{c}_{i,k}$}
            $p \gets \textnormal{BlendColor}(p,\, \hat{\mathbf{x}},\, \boldsymbol{\mu}^{\textnormal{2D}}_{i,j},\, \boldsymbol{\Sigma}^{\textnormal{2D}}_{i,k},\, \boldsymbol{c}_{i,k},\, o_i)$\;
        }
        $\mathbf{I}_{LF}[\hat{\mathbf{x}}] \gets p$\;
        \Return $\mathbf{I}_{LF}$
    }
\end{algorithm}

\begin{table}[t]
    \caption{\textbf{Rendering speed on an RTX 3090.} Even on the cache-limited GPU, \ours{} consistently outperforms the baselines.}
    \label{table:rtx3090}
    
    \resizebox{1.0\linewidth}{!}{
\begin{tabular}{l|cc|cc}
\Xhline{2\arrayrulewidth}
\makecell[c]{\multirow{2}{*}{Method}}     & \multicolumn{2}{c|}{Synthetic Blender} & \multicolumn{2}{c}{Mip-NeRF 360} \\
\cline{2-5}
                            & \multicolumn{1}{c|}{2K FPS}    & 4K FPS        & \multicolumn{1}{c|}{2K FPS}  & 4K FPS    \\ 
\hline
\hline
3DGS                        & \multicolumn{1}{c|}{4.3}   & 2.3       & \multicolumn{1}{c|}{2.3} & 1.1   \\
3DGS (batch=18)             & \multicolumn{1}{c|}{8.5}   & 5.4       & \multicolumn{1}{c|}{3.1} & 1.7   \\
\hline
Ours w/o Reuse w/o Remap    & \multicolumn{1}{c|}{14}    & 8.9       & \multicolumn{1}{c|}{5.3} & 2.9   \\
Ours w/o Reuse              & \multicolumn{1}{c|}{15}    & 10        & \multicolumn{1}{c|}{5.6} & 3.2   \\
Ours w/o Remap              & \multicolumn{1}{c|}{26}    & 16        & \multicolumn{1}{c|}{8.1} & 4.0   \\
Ours                        & \multicolumn{1}{c|}{36}    & 23        & \multicolumn{1}{c|}{12}  & 6.0   \\
\Xhline{2\arrayrulewidth}
\end{tabular}
    }
    
\end{table}

\paragraph{Alpha Blending.}
For each thread with rank $r$ corresponding to an output subpixel, we obtain its remapped coordinate $\hat{\mathbf{x}} = \Psi_{\textnormal{remap}}(r)$, then identify the tile ID $t$, per-subpixel viewpoint index $j = \mathbf{V}[\hat{\mathbf{x}}]$, and cluster ID $k$. We retrieve the sorted Gaussian range $[S_{t,k}, E_{t,k})$ for that (tile, cluster) pair via \texttt{IdentifyGaussianList}, and accumulate the pixel color by front-to-back alpha blending using the per-view $\boldsymbol{\mu}^{\textnormal{2D}}_{i,j}$ alongside the reused cluster-shared $\boldsymbol{\Sigma}^{\textnormal{2D}}_{i,k}$, $\boldsymbol{c}_{i,k}$.

\begin{figure}[t]
    \centering 
    \includegraphics[width=0.8\linewidth]{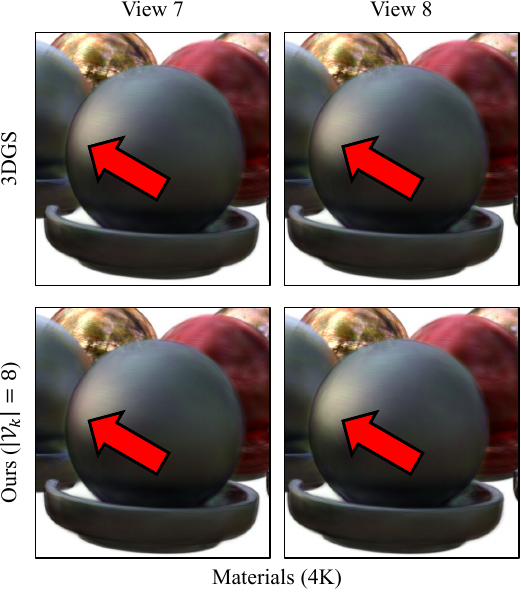}
    \caption{\textbf{Artifacts on specular surfaces.} \viewreuse{} can introduce artifacts on highly specular surfaces (\textit{Materials} scene from the Synthetic Blender dataset)}
    \Description{specular_qualitative}
    \label{fig:specular_qualitative}
\end{figure}

\section{Additional Analysis}
\label{sec:supp_analysis}

\subsection{Evaluation under Limited L2 Cache}
\label{sec:supp_rtx3090}

We evaluate our method on an RTX 3090 with a smaller L2 cache (6MB) than the RTX 5090 (96MB). Due to the 24GB VRAM budget of this GPU, the 3DGS baseline with view batching is limited to a batch size of 18.
As reported in \cref{table:rtx3090}, \ours{} consistently outperforms all baselines even on this cache-limited GPU.

While \viewremapping{} introduces scattered writes when storing final subpixel colors, this occurs only once per subpixel after alpha blending. Conversely, projected Gaussian attributes are read repeatedly within the blending loop. Coalescing these frequent reads outweighs the overhead of the single scattered write. The evaluation results on the RTX 3090 confirm that this method remains effective even with a small L2 cache.

\begin{table*}[t]
    \caption{\textbf{Rendering time breakdown on the Synthetic Blender dataset.} All times are in milliseconds. The number of Gaussian-tile pairs are in millions.}
    \label{table:profiler_blender}
    
    \resizebox{\linewidth}{!}{
\begin{tabular}{l|cccccc|cccccc}
\Xhline{2\arrayrulewidth}
\makecell[c]{\multirow{2}{*}{Method}}                                & \multicolumn{6}{c|}{2K}                                                                                                                                  & \multicolumn{6}{c}{4K}                                                                                                                                    \\
\cline{2-13}
                          & \multicolumn{1}{c|}{Proj.} & \multicolumn{1}{c|}{KeyGen} & \multicolumn{1}{c|}{Sort} & \multicolumn{1}{c|}{Blend} & \multicolumn{1}{c|}{VRAM} & \#Pairs & \multicolumn{1}{c|}{Proj.} & \multicolumn{1}{c|}{KeyGen} & \multicolumn{1}{c|}{Sort} & \multicolumn{1}{c|}{Blend} & \multicolumn{1}{c|}{VRAM} & \#Pairs \\ 
\hline
\hline
Ours w/o Reuse w/o Remap        & \multicolumn{1}{c|}{0.76}  & \multicolumn{1}{c|}{1.07}   & \multicolumn{1}{c|}{0.14} & \multicolumn{1}{c|}{15.73} & \multicolumn{1}{c|}{4.22GB} & 76.0M   & \multicolumn{1}{c|}{0.83}  & \multicolumn{1}{c|}{1.47}   & \multicolumn{1}{c|}{0.14} & \multicolumn{1}{c|}{21.27} & \multicolumn{1}{c|}{6.25GB} & 134.3M  \\
Ours w/o Reuse                  & \multicolumn{1}{c|}{0.75}  & \multicolumn{1}{c|}{1.06}   & \multicolumn{1}{c|}{0.14} & \multicolumn{1}{c|}{7.90}  & \multicolumn{1}{c|}{4.25GB} & 76.0M   & \multicolumn{1}{c|}{0.83}  & \multicolumn{1}{c|}{1.48}   & \multicolumn{1}{c|}{0.14} & \multicolumn{1}{c|}{10.67} & \multicolumn{1}{c|}{5.89GB} & 134.3M  \\
Ours w/o Remap                  & \multicolumn{1}{c|}{0.18}  & \multicolumn{1}{c|}{0.57}   & \multicolumn{1}{c|}{0.13} & \multicolumn{1}{c|}{10.00} & \multicolumn{1}{c|}{2.77GB} & 15.6M   & \multicolumn{1}{c|}{0.20}  & \multicolumn{1}{c|}{0.91}   & \multicolumn{1}{c|}{0.13} & \multicolumn{1}{c|}{17.09} & \multicolumn{1}{c|}{3.54GB} & 27.5M   \\
Ours                            & \multicolumn{1}{c|}{0.18}  & \multicolumn{1}{c|}{0.56}   & \multicolumn{1}{c|}{0.13} & \multicolumn{1}{c|}{4.45}  & \multicolumn{1}{c|}{3.01GB} & 15.6M   & \multicolumn{1}{c|}{0.20}  & \multicolumn{1}{c|}{0.91}   & \multicolumn{1}{c|}{0.13} & \multicolumn{1}{c|}{10.41} & \multicolumn{1}{c|}{4.09GB} & 27.5M   \\
\Xhline{2\arrayrulewidth}
\end{tabular}
    }
    
\end{table*}

\begin{table*}[t]
    \caption{\textbf{Rendering time breakdown on the Mip-NeRF\,360 dataset.} All times are in milliseconds. The number of Gaussian-tile pairs are in millions.}
    \label{table:profiler_mipnerf}
    
    \resizebox{\linewidth}{!}{
\begin{tabular}{l|cccccc|cccccc}
\Xhline{2\arrayrulewidth}
\makecell[c]{\multirow{2}{*}{Method}}                                & \multicolumn{6}{c|}{2K}                                                                                                                                    & \multicolumn{6}{c}{4K}                                                                                                                                      \\
\cline{2-13}
                          & \multicolumn{1}{c|}{Proj.} & \multicolumn{1}{c|}{KeyGen} & \multicolumn{1}{c|}{Sort} & \multicolumn{1}{c|}{Blend} & \multicolumn{1}{c|}{VRAM}   & \#Pairs & \multicolumn{1}{c|}{Proj.} & \multicolumn{1}{c|}{KeyGen} & \multicolumn{1}{c|}{Sort} & \multicolumn{1}{c|}{Blend} & \multicolumn{1}{c|}{VRAM}   & \#Pairs \\ 
\hline
\hline
Ours w/o Reuse w/o Remap        & \multicolumn{1}{c|}{1.63}  & \multicolumn{1}{c|}{5.24}   & \multicolumn{1}{c|}{0.18} & \multicolumn{1}{c|}{33.87} & \multicolumn{1}{c|}{10.18GB} & 259.6M  & \multicolumn{1}{c|}{1.92}  & \multicolumn{1}{c|}{9.04}   & \multicolumn{1}{c|}{0.26} & \multicolumn{1}{c|}{67.43} & \multicolumn{1}{c|}{18.12GB} & 551.5M  \\
Ours w/o Reuse                  & \multicolumn{1}{c|}{1.63}  & \multicolumn{1}{c|}{5.26}   & \multicolumn{1}{c|}{0.18} & \multicolumn{1}{c|}{24.13} & \multicolumn{1}{c|}{10.20GB} & 259.6M  & \multicolumn{1}{c|}{1.92}  & \multicolumn{1}{c|}{9.05}   & \multicolumn{1}{c|}{0.26} & \multicolumn{1}{c|}{48.29} & \multicolumn{1}{c|}{18.12GB} & 551.5M  \\
Ours w/o Remap                  & \multicolumn{1}{c|}{0.31}  & \multicolumn{1}{c|}{3.09}   & \multicolumn{1}{c|}{0.14} & \multicolumn{1}{c|}{31.07} & \multicolumn{1}{c|}{2.85GB}  & 62.8M   & \multicolumn{1}{c|}{0.34}  & \multicolumn{1}{c|}{4.87}   & \multicolumn{1}{c|}{0.15} & \multicolumn{1}{c|}{61.03} & \multicolumn{1}{c|}{5.01GB}  & 129.7M  \\
Ours                            & \multicolumn{1}{c|}{0.30}  & \multicolumn{1}{c|}{3.08}   & \multicolumn{1}{c|}{0.14} & \multicolumn{1}{c|}{18.28} & \multicolumn{1}{c|}{2.85GB}  & 62.8M   & \multicolumn{1}{c|}{0.35}  & \multicolumn{1}{c|}{4.98}   & \multicolumn{1}{c|}{0.15} & \multicolumn{1}{c|}{32.41} & \multicolumn{1}{c|}{5.01GB}  & 129.7M  \\
\Xhline{2\arrayrulewidth}
\end{tabular}
    }
    
\end{table*}

\subsection{Rendering Time Breakdown}
\label{sec:supp_profiler}

To analyze the performance gains of \ours{}, we report the per-stage execution time, peak VRAM usage, and the total number of Gaussian-tile pairs (\#Pairs). The evaluated stages include projection (Proj.), key generation (KeyGen), sorting (Sort), and alpha blending (Blend).

\Cref{table:profiler_blender,table:profiler_mipnerf}  detail the results on the Synthetic Blender and Mip-NeRF 360 datasets, respectively. \viewreuse{} reduces projection, key generation, and sorting times by eliminating redundant computations. \viewremapping{} decreases blending time by enabling coalesced memory accesses. When combined, these two components further reduce rendering time and enable interactive frame rates while lowering memory consumption.

\subsection{Impact of View-Coherent Attribute Reuse on Rendering Quality}
\label{sec:supp_specular}

\viewreuse{} can introduce artifacts on highly specular surfaces. In practice, commercial LFDs sample viewpoints with high angular density, which keeps the angular deviation within each cluster very small. Furthermore, the spherical harmonics used in 3DGS naturally attenuate the high-frequency components of specular reflections. Although specular inconsistencies are observable under extreme conditions (see \cref{fig:specular_qualitative}), our qualitative evaluations and ablations demonstrate that reusing attributes does not lead to significant perceptual degradation.

\end{document}